%% file: main.tex
\documentclass[twocolumn,10pt,numbers]{sigplanconf}

\usepackage{graphicx}

\usepackage{amsfonts}
\usepackage{comment}
\usepackage{times}
\usepackage{amsmath}
\usepackage{changepage}
\usepackage{amssymb}

\usepackage{listings}

\usepackage[hyphens]{url}

\usepackage{datetime}

\usepackage{algorithmic}
\usepackage[linesnumbered,ruled]{algorithm2e}

\newcommand\textvtt[1]{{\normalfont\fontfamily{cmvtt}\selectfont #1}}

\begin{document}


\input{./subsection/000_title.tex}

\input{./subsection/010_abstract.tex}

\section{Introduction}\label{S_introduction}
\input{./subsection/020_introduction.tex}

\section{Background and Related Work}\label{S_background_n_relatedwork}
\input{./subsection/040_background_related.tex}

\section{Design and Implementation}\label{S_design}
\input{./subsection/050_design_implement.tex}
\section{Evaluation}\label{S_evaluation}

\input{./subsection/060_evaluation.tex}

\section{Discussion}\label{S_discussion}

\input{./subsection/070_discussion.tex}

\section{Conclusion}\label{S_conclusion}
\input{./subsection/080_conclusion.tex}


\acks
\input{./subsection/090_acknowledgment.tex}

\input{./subsection/100_reference.tex}

\end{document}

%% file: subsection/000_title.tex
\publicationrights{licensed}

\setlength{\pdfpageheight}{\paperheight}
\setlength{\pdfpagewidth}{\paperwidth}

%
%

\copyrightyear{2016} 
\conferenceinfo{EuroSys '16,}{April 18-21, 2016, London, United Kingdom}
\copyrightdata{978-1-4503-4240-7/16/04}
\copyrightdoi{2901318.2901320}

\title{BB: Booting Booster for\\Consumer Electronics with Modern OS}
%
\authorinfo{Geunsik Lim \and MyungJoo Ham}
			{Software R\&D Center, Samsung Electronics}
			{\{geunsik.lim, myungjoo.ham\}@samsung.com}
%
%


%
%

\maketitle

%% file: subsection/010_abstract.tex
%
\begin{abstract}
%

Unconventional computing platforms have spread widely and rapidly following smart phones and tablets: consumer electronics such as smart TVs and digital cameras. For such devices, fast booting is a critical requirement; waiting tens of seconds for a TV or a camera to boot up is not acceptable, unlike a PC or smart phone. Moreover, the software platforms of these devices have become as rich as conventional computing devices to provide comparable services. As a result, the booting procedure to start every required OS service, hardware component, and application, the quantity of which is ever increasing, may take unbearable time for most consumers. To accelerate booting, this paper introduces \textit{Booting Booster} (BB), which is used in all 2015 Samsung Smart TV models, and which runs the Linux-based Tizen OS. BB addresses the init scheme of Linux, which launches initial user-space OS services and applications and manages the life cycles of all user processes, by identifying and isolating booting-critical tasks, deferring non-critical tasks, and enabling execution of more tasks in parallel. BB has been successfully deployed in Samsung Smart TV 2015 models achieving a cold boot in 3.5 s (compared to 8.1 s with full commercial-grade optimizations without BB) without the need for suspend-to-RAM or hibernation. After this successful deployment, we have released the source code via \url{http://opensource.samsung.com/}, and BB will be included in the open-source OS, Tizen (\url{http://tizen.org/}).

\end{abstract}


%% file: subsection/020_introduction.tex
Consumers desire a shorter response time to user input for consumer electronics, including TVs. This is usually achieved using newer and faster hardware. An exception is the booting time of TVs. Decades ago, the picture could be seen and the channel changed within a  few seconds after turning on: 5 s in 1974 \cite{instant_warm_cathode} and  $\sim$1~s in 1994 \cite{fast_warmup_picture,quick_heating_cathode}. Today, many seconds are required with modern smart TVs. Such degeneration is usually incurred by the increased number and complexity of software packages. Consumer electronic devices are supposed to respond quickly to users and not always be on--e.g., TVs and cameras--booting time is a critical performance metric. For TVs, users want to see the picture and change channels immediately. With digital cameras, users want to take photographs immediately; otherwise, users may miss the scene. We describe in this work a significant reduction in the booting time of recent smart TVs running the Linux-based Tizen OS \cite{tizen_association}.

To reduce the booting time of consumer electronics with general-purpose operating systems (a.k.a. smart devices), developers have used various approaches, including hibernation \cite{kaminaga2006improving} and suspend-to-RAM \cite{jo09_improving}. With a continuous power supply, a device may use suspend-to-RAM, which stores all hardware states into RAM and keeps RAM powered while users consider the device to be turned off (most smart phones do this). However, many consumers are sufficiently sensitive to energy consumption to unplug TVs while not in use (terminating suspend-to-RAM), and complain of subsequent slower booting. We have been using suspend-to-RAM approaches for high-end TVs to meet the booting time requirement and have received such complaints.


%

Another popular approach is hibernation, which stores all states, including volatile memory (DRAM) contents in non-volatile memory (flash memory and hard disk drives) and restores the states upon booting \cite{kaminaga2006improving}. We have been applying such methods to digital cameras with Tizen, and achieved a booting time of less than 2 s for the NX-300, which does not support third-party applications. If the initial states after booting are not constant due to third-party applications and user customization, the stored hibernation images must be updated, which may trigger a critical issue. Updating hibernation images require excessive time for device power off, and so users cannot turn the device on or unplug it immediately after turning it off. Thus, for smart TVs, cold boot (conventional booting) performance is critical.

%

Smart TVs usually suffer from longer booting times. For example, some smart TVs take 20 s \cite{boot_time_sonytv_20s} or even 40 s \cite{boot_time_hisense_40s} to boot. This time was less than 2 s decades ago! As mentioned above, users do unplug TVs; thus, suspend-to-RAM works in limited cases only, although suspend-to-RAM is extremely effective; e.g., less than 2 s with a suspend-to-RAM based ``Instant-On'' function \cite{instant_on_samsung_2s}. Therefore, major performance goals for latest Tizen TVs include a faster cold boot time: 3.5 s \footnote{Natural interactions between human and computer appear to require a response time of 2.3 to 3.5 s \cite{miller1968response}.} after plugging in and turning on a smart TV.


This paper shows how system software developers have achieved major breakthroughs in booting performance without understanding most software packages in the system, which are too vast and too dynamically changing to be followed by a small number of system software developers (three in our case). A cold boot starts with boot loaders, which load a kernel after completion. At the completion of kernel booting, the kernel loads an init scheme \cite{Royon07_CoRR_init-survey}, which loads and initializes predefined software packages, including OS services and start-up applications. With continuous increases in the number of peripheral hardware components and the number of OS services, the task size of booting has increased continuously for both kernel and user spaces. To cope with the ever-increasing number of hardware and software components to be initialized for booting, developers have attempted to exploit parallelism with the multi-core CPUs is available to recent consumer electronics. For example, Samsung JS9500 series TVs have eight CPU cores.

%
%

Unfortunately, tasks in booting sequences are not independent of each other and the dependencies between tasks may become extremely complex. The complexity of dependencies is especially problematic with OS services, where both the number of services and the number of dependency relations between them \cite{Nakamura09_APSCC_service_candidates} may increase significantly during development. We have witnessed a case in which the number of OS services started at slightly over 100 and doubled in a few months. Another aspect is the involvement of many developers in such a large software project, most of who have a limited understanding of the OS, including the init scheme. As a result, to optimize or stabilize their own software component, developers often declare dependencies and orderings excessively and unnecessarily. Note that many different types of dependencies may be declared with modern Linux init schemes; e.g., ``I need A'', ``I am needed by B'', ``I do not want to be launched after C'', and ``I want to be launched after file path D is available'' \cite{systemd_official_freedesktop}. At the same time, the number of components and the relationships between the components of the software platform are too large for the few system software developers (the authors) to address directly, particularly when the platform is dynamically changed by fellow developers on a daily, and indeed hourly, basis.

%
%

This paper describes how we have reduced significantly the cold boot time of smart TVs with numerous software components and complex dependencies without working on such software components directly. The proposed mechanism is general enough for other Linux platforms so long as they run a recent version of the de facto standard init scheme, \textit{systemd}, (v208 or later) and a recent Linux kernel version (3.10 or later). Our main contributions include:
%

%
%

\begin{itemize}
\item We developed the \textit{Booting Booster} (BB) mechanism, which quickly initializes the system and launches crucial start-up applications and their services significantly earlier by:
\begin{itemize}
	\item Deferring tasks that are not crucial determinants of the booting time.
    \item Automatically identifying, isolating, and prioritizing tasks critical to booting completion.
    \item Adopting a booting time-aware synchronization mechanism and improved parallelism with more tasks executed in parallel (modularize and defer).
    \item Pre-processing repeated tasks, such as loading and parsing of service configuration files, at build-time.
    \item Improving the performance of bottlenecks in the infrastructure.
\end{itemize}


%
%

\item We have successfully deployed BB and its supporting mechanisms for all 2015 models of Samsung Smart TVs globally after demonstrating improved performance with some 2014 models as described in \S\ref{S_evaluation}. The source code is available to the general public at \url{http://opensource.samsung.com/}, and will be included in the later versions of the open-source OS, Tizen, available at \url{http://tizen.org/}.
\end{itemize}

%

The rest of this paper is organized as follows. The next section, \S\ref{S_background_n_relatedwork}, describes in more detail the background and previous work on reducing booting time, together with the need for the traditional cold boot for consumer electronics with modern OSs. \S\ref{S_design} describes how the proposed mechanism, BB, is implemented. \S\ref{S_evaluation} shows the evaluation methods and experimental results of BB. We further discuss the issues of booting and BB in \S\ref{S_discussion}. Finally, we conclude the paper in \S\ref{S_conclusion}.

%% file: subsection/040_background_related.tex


\begin{figure}
\centering
\includegraphics[width=0.99\columnwidth,height=1.0in]{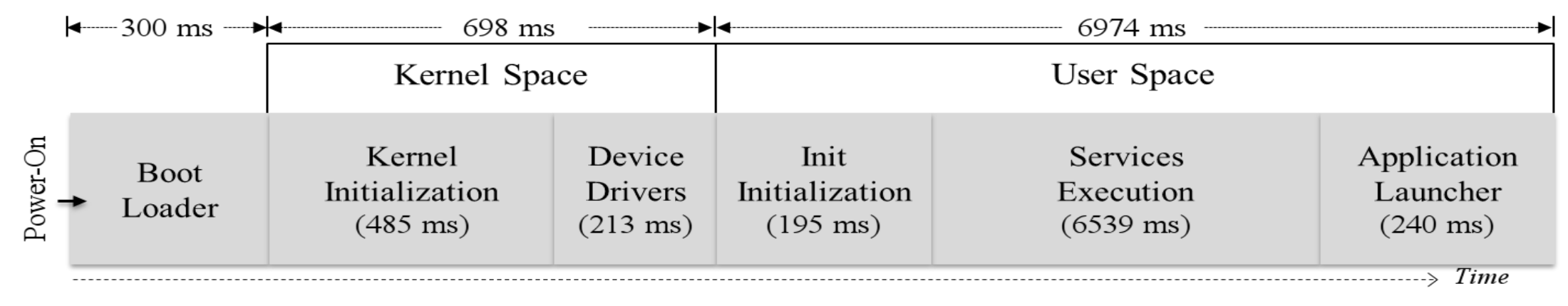}
\caption{Overall booting sequence of a TV}
\label{fig:overall_booting_time_before}
\end{figure}

%

Booting modern consumer electronics based on Linux is in general not different from booting a Linux desktop computer. Figure~\ref{fig:overall_booting_time_before} shows the overall booting procedure with timing information of a Tizen TV before this work. Note that optimization techniques of prior work and commercial-level optimization of software packages are already applied to the booting procedure shown in Figure~\ref{fig:overall_booting_time_before}.

%

Upon receiving a turn-on signal, the CPU initiates the booting procedure using instructions stored in the internal read-only memory (ROM). These instructions cause the CPU to load another set of instructions (a bootloader) from a predefined location in a storage device and launch the bootloader. The bootloader initiates the hardware components required to start the kernel, and loads and launches the kernel.

%
After the kernel has been initialized (\S\ref{SS_background_kernel})--i.e., all kernel components and device drivers have started and are ready to start user-space components--the kernel invokes an init process~\cite{init_definition}, which is the first user process and runs as a daemon until the system shuts down (\S\ref{SS_background_init}). Without the init process, the kernel halts the system with a kernel panic. The init process takes charge of user process management, including boot-up and shut-down sequences. Therefore, optimizing the init process and the kernel is the starting point for reducing booting time, especially if the init process is complex and heavy, as many modern init schemes are.

%
%

The definition of boot completion may differ by device type. For TVs, we define booting completion if 1) the video and audio of a broadcast channel is played and 2) it responds to remote control inputs; i.e., a user may change channels and see and hear the selected channel. For cameras, booting is completed if lenses and sensors are ready to capture the scene and the display is showing what the lenses are seeing. Phones are often considered to have booted if the user can make a phone call.

\subsection{Suspend and Restore}

%
We have applied two suspend-and-restore approaches in consumer electronics: hibernation-based snapshot booting \cite{jo09_improving,kaminaga2006improving} and suspend-to-RAM \cite{jo09_improving}. Suspend-and-restore approaches allow the system to skip most booting tasks. Restore mechanisms of hibernation may be implemented in a bootloader (e.g., Samsung NX-300M). Restore mechanisms of suspend-to-RAM are also often implemented in the internal ROM of application processors to skip bootloaders (e.g., the recent Samsung Exynos series).

%

For traditional consumer electronics lacking third-party applications or application marketplaces, snapshot booting has been effective even for Linux-based devices. For example, NX-300M cameras \cite{opensource_samsung} that run the Linux-based Tizen OS (without the Tizen store) have achieved a $\sim$1~s booting time \cite{nx300_1.2sec_stratup} with snapshot booting. Such devices have limited variations in their initial states after booting, which in turn, enables use of pre-loaded factory snapshot images. If users may install third-party applications and services, as allowed by smart phones and smart TVs, pre-loaded snapshot images become useless and the device must create snapshot images at run-time. However, users may unplug power cords or pull out batteries at any time (disrupting the image creation process), and creating a snapshot image takes much time, during which users cannot use the device. Forcing users to wait for tens of seconds (if not over 1 minute) with no response during application installation or shutdown is even worse than slow booting of tens of seconds~\cite{kaminaga2006improving}. Moreover, with larger DRAM size, snapshot booting may take a lot of time. For example, the universal flash storage (UFS) 2.0 internal flash media of Galaxy S6, which is one of the most advanced storage devices for mobile phones, can read sequentially at $\sim$300~MiB/s \cite{galaxy_s6_ufs_benchmark}, which means that 10 s are required to read 3~GiB (the DRAM size of the Galaxy S6).

%

Suspend-to-RAM has been widely applied to consumer electronics: smart phones, TVs (Instant On \cite{kaminaga2006improving}), cameras (NX300M uses both snapshot booting and suspend-to-RAM), and watches. A critical requirement of suspend-to-RAM is supply of power to the device while suspended. For battery-operated devices, this might not be an issue. However, many users unplug TVs frequently, prohibiting suspend-to-RAM, and complain of slow booting. Putting batteries into TVs incurs additional manufacturing costs and requires TV developers to optimize power consumption so that TVs may stay ``turned off'' for longer. Another issue is that some hardware components of budget TV models do not support suspend-to-RAM, but still run a general-purpose OS (Tizen), while the booting time requirement is not much different. Thus, we cannot rely on suspend-to-RAM for TVs (and other plugged-in consumer electronics), and so we should make a cold boot--booting up the device with empty RAM and conventional booting sequences--sufficiently fast.


Another suspend-to-RAM based approach is to immediately boot up silently (keeping the screen off) if a TV is plugged in and suspend the TV after booting until a user presses the power button. Unfortunately, this idea was rejected because such behaviors may violate a regulation of the European Union \cite{european_union_regulation_standby}. According to this regulation, the power consumption of a TV in standby cannot exceed 1 W. An active smart TV application processor consumes well over 1~W.

\subsection{Non-Volatile RAM (NVRAM) Based Approaches}



Tim Bird \cite{bird04_methods} has suggested applying eXecution-In-Place (XIP) \cite{joo06_demand} to non-volatile RAM (NVRAM) to accelerate booting. The system instantaneously becomes ready from power-off with NVRAM because the states are preserved during power-off.

The introduction of NVRAM as fast as DRAM, such as MRAM \cite{nvram_tehrani99_mram}, PRAM \cite{nvram_oh06_512mb_pram}, and RRAM \cite{nvram_zhuang02_rram}, may enable this approach. However, high-density and high-performing NVRAM is too expensive for mass production \cite{where-nvram-fit}. Affordable high-performance NVRAM provides only tens of MiBs at this point; 16~MiB MRAM costs twenties of US dollars \cite{everspin_mram_low_cost}. Intel and Micron have announced their mass production plan for 3D CrossPoint memory \cite{micron_3d_xpoint}, which is claimed to be byte-addressable, non-volatile, high-density, and faster than conventional flash media, but still slower than DRAM \cite{idf2015}. As no affordable high-density and high-performing (at least as DRAM) NVRAM is available now and for the foreseeable future, we ignore NVRAM technology in this study.

\subsection{Compression}\label{SS_background_compression}


The I/O throughput of flash memory had been the major bottleneck of booting of consumer electronics. Thus, compression had been widely accepted to accelerate booting. However, compression is of little help because the flash I/O throughput \cite{ken12_flash,inconvenient_truth_of_nand,emmc_to_ufs,kozuch94_compression_of_embedded} exceeds decompression throughput. With the Galaxy S6, running all eight cores provides 35~MiB/s decompression throughput \cite{galaxy_s6_geekbench}, while the embedded flash storage provides 300~MiB/s sequential read throughput \cite{galaxy_s6_ufs_benchmark}.

\subsection{Kernel Booting}\label{SS_background_kernel}



The complexity of the kernel itself is the main barrier to fast kernel booting: the number of devices, their drivers, and kernel subsystems. The size of the kernel binary (10 MiB) is not a major concern for booting time. Kernel complexity has increased continuously with the evolution of software platforms for modern consumer electronics (a.k.a. smart devices) and the tendency to embed more peripheral devices. For example, a 2015 model of Samsung Smart TV has 408 kernel modules (.ko files) \cite{build-external-module}.

We have also applied conventional optimization methods for the kernel \cite{kaminaga2006improving,asberg13_fast,debugging-by-printing,bovet2005understanding,krishnakumar2005kernel}. Along with such optimizations, we have identified and disabled unnecessary kernel components, such as debugging, tracing, logging, and profiling mechanisms, as well as extensive kernel modualization to defer a significant portion of the kernel initialization. Such improvements have reduced the kernel booting time from 6.127 s to 0.698 s, which is the base performance before application of the proposed BB mechanism.

\begin{figure*}
\centering
\includegraphics[width=0.99\textwidth,height=3.5in]{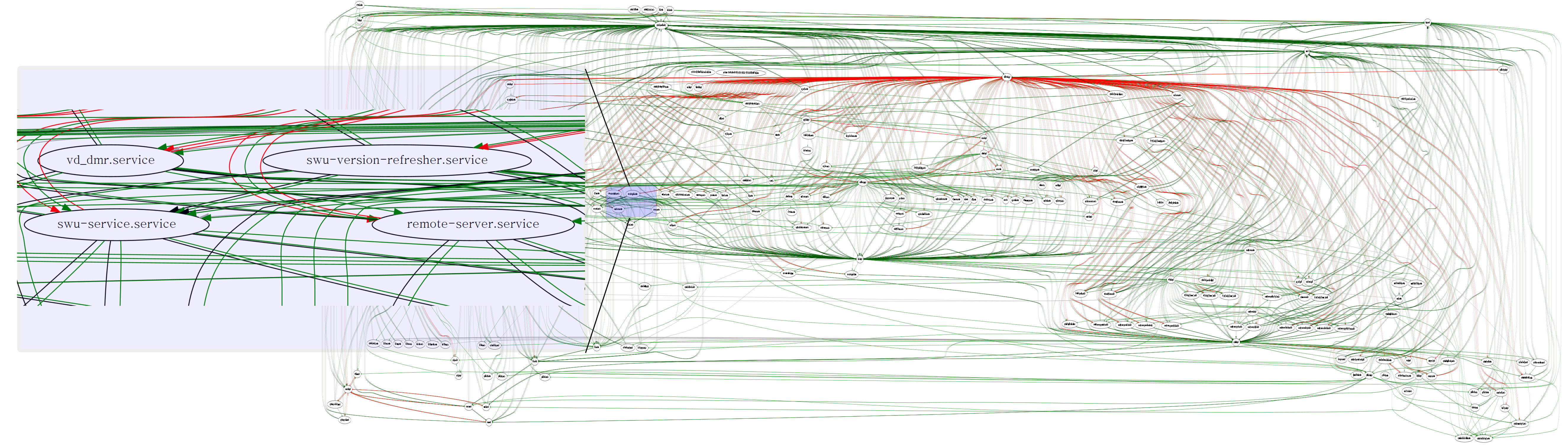}
\caption[]{Dependencies between 136 services of the open-source Tizen TV OS. Red lines are strong dependencies (launch B after A is ready) and green lines are weak dependencies (launch B not before launching A). Additional dependency types are supported by \textit{systemd}, and shown in other colors \cite{systemd_unit}. For commercialization, the number of nodes has almost doubled within a few months.}
\label{fig:dependency-complexity}
\end{figure*}

\subsection{Booting of User Space with Init Schemes}\label{SS_background_init}

%
%
%
%
%

Figure~\ref{fig:dependency-complexity} shows the dependency relations among 136 services of Tizen TV handled by the init scheme, \textit{systemd}, for booting. Moreover, a service may include multiple processes (usually about three); e.g., the D-Bus IPC service has three processes. In a fork for a product, the number of the services has increased to more than 250 from 136 in a few months. This is because of numerous additional requirements from consumers, content providers, network operators, and manufacturers of the “mainline” OS targeting general devices and vendors. For example, some content providers require their own digital rights management (DRM) solutions to be embedded, which are not approved to be open sourced; therefore, open-source communities do not accept such components \cite{fsf_opposing_drm}.


The complexity of dependencies is a major barrier to booting time optimization. If a system administrator changes the start time of a service to start the service earlier by adjusting priorities or dependencies, the start time of other services may be changed significantly, resulting in unexpected side-effects on the overall boot time. Another problem with such complexity is non-deterministic behavior of booting sequences. Many paths exist in booting sequences, with numerous services to be initialized and sparse dependency relations. Indeed, the initialization time of a service may be not constant, especially if it depends on network responses or user input. The most serious headache of system administrators for boot-time optimization is that the services and their relations themselves change dynamically version by version. The administrators are not dealing with static software components, but with components continuously updated by fellow developers, who also have issues with commercialization.


%

Init schemes have evolved continuously since the appearance of BSD init \cite{mewburn01_design} to initialize OS services correctly at boot time \cite{systemd_official_freedesktop}. The first widely adopted init scheme after BSD init, rcS \cite{linux-start-scripts,creating-rcs-script}, had no parallelism; i.e., only one service is initiated at a time \cite{mewburn01_design}. Linux start-up script (e.g., /etc/init.d/rcS) allows the init daemon to run additional programs at boot time. Its typical use is to mount additional filesystems and launch daemons. With the introduction of multi-processor systems, rcS evolved to support parallelism by adopting  \textvtt{fork} and \textvtt{execve} system calls \cite{Royon07_CoRR_init-survey}. Later, rcS adopted a multi-threaded structure with the POSIX thread library, enabling multi-threaded programs in rcS script to be more efficient while supporting parallelism.


%

Init schemes have recently started managing CPU scheduling, I/O scheduling, memory pressure, as well as the creation and termination of the process in user-space. Such functionality has been traditionally managed solely by the kernel. However, the kernel usually treats every user process equally and does not have enough information to treat each user process according to its specifications in general. To fill such gaps, recent init schemes have added resource management mechanisms \cite{adams05_solaris} with policies based on the characteristics of each registered service and user processes in general.

\begin{itemize} 
\item \textbf{CPU scheduling} sets the default scheduling priority 
for executed processes with the POSIX system calls such as \textvtt{nice()}, 
\textvtt{setpriority()}, and \textvtt{sched\_setscheduler()}.
\item \textbf{I/O scheduling} sets the I/O scheduling class and the I/O scheduling 
priority for running processes with the \textvtt{ioprio\_set()} system call.
\item \textbf{Memory pressure} management adjusts priorities between
user processes and chooses the victim to be expelled from the main
memory when the memory pressure becomes critical.
\item \textbf{User-process creation and termination} capability enables 
the init scheme to manage the life cycle of all user processes.
\end{itemize}

As Figure~\ref{fig:overall_booting_time_before} suggests, the time required to start OS user-space services and initiate device drivers takes most of the booting time after optimization, based on prior work. The growing complexity of the software platform and the growing number of peripheral devices greatly contribute to the increased time required for such activities. In addition, as described in the previous paragraph, the growing complexity of required services from the init scheme itself for user space processes contribute to the delay in boot time. This is because the modern init scheme is required to take account of all dependency relationships in addition to the start-up orders, and the start time of the services is not determined at build time; i.e., users may install additional services, services may be updated (network operators and manufacturers may update services without the user's knowledge), or a service may update its own description at any time.



\subsubsection{Out-of-Order Mechanisms}


Out-of-order mechanisms execute a service without consideration of completion of services intended to be prior to the service: BSDinit, SysVinit, eINIT, Launchd, Svscan, Windows service control manager, and Busybox-init \cite{mewburn01_design,sysvinit,einit,launchd,svscan,win-service-controller,busybox-init}. In other words, any service may instantly start at a specified time due to out-of-order mechanisms. The major drawback of out-of-order mechanisms is that they cannot handle the boot sequence correctly if the dependency relationship and start time of a service are changeable at run time (i.e., not determined at build time) \cite{debian_fork_devuan}, the service start up latency is not deterministic, or relations between services or services themselves are being changed. As mentioned above, all such adverse conditions hold for modern consumer electronics.

To mitigate such drawbacks, out-of-order mechanisms have recently adopted a path-check method that delays a service start-up until the creation of the specified path, which is created by another service required by the service. As a result, recent out-of-order mechanisms have become partially in-order for specified and modified services. However, such methods require modifications in all OS services that have the possibility of non-determinism and dependency relations, which is not feasible if the dependency relations are too complex to be handled manually with custom paths for each dependency, as in our systems.

%
%
\subsubsection{In-Order Mechanisms}

%

In-order mechanisms,  \textit{Advanced Boot Script},  \textit{OpenRC},  \textit{Upstart}, and \textit{systemd}, mandate the completion of launching of required services before launching a dependent service. In-order mechanisms guarantee the correct start-up sequence of services while in parallel, invoking non-interdependent user processes. In-order mechanisms completely eliminate the possibility of incorrect booting sequences due to the non-determinism and dynamicity so long as each service has fully described services upon which it depends.

Advanced Boot Script \cite{gooch02_advanced} has proposed an in-order init scheme supporting parallelism and dependency declaration. Advanced Boot Script allows software package developers to declare uni-directional dependencies (e.g., \textit{``I need service A initialized before me''}) for services, and initializes services in parallel if they are in a same group and do not break the declared dependencies. The limitations of Advanced Boot Script for boot time optimization are as follows. 1) It is based on run-levels, which are groups subscribed by each program to be invoked at boot-time, and run-levels are in a total order. Programs in different run-levels cannot be invoked in parallel. 2) It does not allow system developers (or administrators in server systems) to prioritize specific programs for faster booting \cite{asberg13_fast}. Note that Advanced Boot Script has focused on guaranteeing the correctness of booting sequences defined by dependency declarations. It does not allow launching tasks of run-levels in parallel or expressing priorities between services.


A more advanced in-order init scheme, \textit{systemd} \cite{systemd_official_freedesktop} has been widely accepted as the standard init scheme for Tizen and other major Linux distributions \cite{systemd_wiki_adoption}: Fedora, Red Hat Enterprise Linux, CentOS, Debian, Ubuntu, Megeia, Arch Linux, CoreOS, Slackware, SUSE Linux Enterprise, and openSUSE. Dependencies, start-up priorities and conditions, resource management policies, monitoring and recovery systems, and other related mechanisms for each service are declared by the service in its own unit file, which is parsed and served at boot time.

%

The current de facto standard Linux init scheme, \textit{systemd}, removes run-levels, which enables execution of more tasks in parallel. \textit{systemd} (officially, it starts with a lowercase ``s'') also allows developers to express more complex service requirements. For example, a service may declare \textit{``I am needed by service A.''}, \textit{``I need to be started before service B''}, or  \textit{``I need to be started if} \textvtt{/tmp/foo}\textit{ is created.''} \textit{systemd} also provides mechanisms to monitor and manage the services at run time and to provide resource management services to user processes. There is much potential to tweak and fine-tune the system using various tools for system and service developers.

\subsubsection{Issues of Modern Init Schemes}

%

To shorten the boot time, we try to execute many processes simultaneously, which exploit the parallelism \cite{edler88_process} supported by multi-core architectures, and aim to fully utilize each core even with I/O-intensive tasks. The major barrier to invoking as many processes as possible at boot time is the possibility of other prerequisite processes not being ready to provide essential services to the invoked processes; if either risks ruining the boot sequence by ignoring the possibilities, or takes more time to guarantee a correct boot sequence, it is very difficult to improve the boot-time of the consumer electronic devices. Figure~\ref{fig:dependency-complexity} shows such dependencies of Tizen OS before the commercialization process, which virtually doubles the number of services (nodes). Init schemes are required to guarantee the correct booting sequence by following the dependencies.

%
\begin{figure*}
\centering
\includegraphics[width=2.0\columnwidth,height=1.9in]{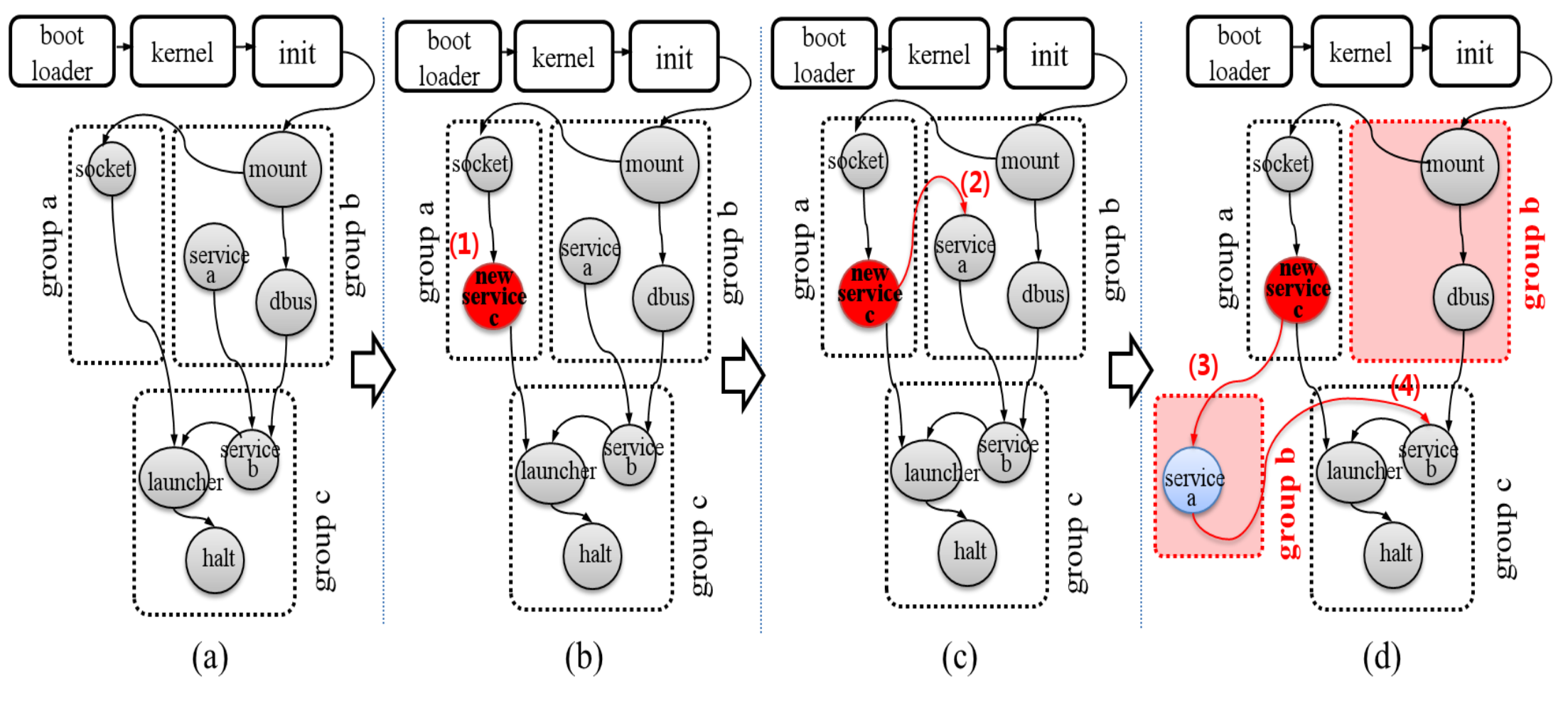}
\caption{Increased complexity of dependency relations by adding a new service; a new service c introduces a cycle between group a and group b, and that cycle forces group b to be split.}
\label{fig:dependency-by-new-service}
\end{figure*}


As such dependencies become more complicated, optimizing the booting sequence \cite{bird04_methods} gets extremely difficult because a single dependency relation may directly affect the start-up times of two services and indirectly affect many others. Figure~\ref{fig:dependency-by-new-service} shows how a single update of a dependency relation may affect overall boot sequences. In the figure, a group represents services to be launched at booting that have similar characteristics and launching orders in the OS and are usually handled by a team of developers. In the figure, a newly added \textvtt{service c} in \textvtt{group a}, which is required by \textvtt{service a} in \textvtt{group b} (or \textvtt{service c} declares that it has to be executed earlier than \textvtt{service a}), affects the whole \textvtt{group c} and partitions \textvtt{group b}, reducing the probability of starting processes in parallel.

Figure~\ref{fig:dependency-by-new-service} suggests that services of a group aligned to a team of developers cannot be guaranteed to be kept as a group (or in a close time line). Then, the corresponding developer team cannot manage the booting sequence of their own services efficiently; their service sequences are easily disrupted by another team. As the number of such exceptions grows (with over 250 OS services, varying pre-loaded applications, varying models per region, operator, display panel size, or market segment, and a lot of developers, it is guaranteed to grow rapidly), a group of services become fragmented and the corresponding developers cannot guarantee the start time of their own services.
In the development of television sets, we have witnessed that one of the most important OS services, D-Bus (the standard inter-process communication (IPC) service of Tizen) daemon \cite{dbus}, suffers from varying start-up time due to the changing requirements and specifications of other services. We describe such behavior in more detail in \S\ref{SS_eval_isolation}.

%
%

Additional drawbacks of \textit{systemd} and other modern in-order mechanisms are due to the complexity of the init scheme itself and the dynamicity, which worsens the effect described in the previous paragraph. Providing more tools and expression capabilities to developers usually results in more chances for abuse. We have witnessed many cases in which service developers attempt to get higher priorities and more resources than others to improve their services, negatively impacting the overall performance; i.e., the booting time. They also often add unnecessary dependencies to ensure the correctness of their services, which results in unnecessarily invoking processes or even creating circular dependencies. With over 250 OS services running in the initial state in smart TVs, system developers do not have sufficient knowledge. Thus, system developers cannot cope with such issues effectively, especially when service developers change specifications day by day until the products are shipped. Sometimes, they change specifications even after the products are shipped!

%
%

On the other hand, executing multiple tasks (initiating OS services) in parallel with complex dependency relations itself is an extremely difficult task that must also be optimized. Besides, reducing conventional (cold) booting time of computer systems down to a few seconds has received little attention because the current mainstream topics in OSs usually focus on the computer systems that do not require a short booting time or may use alternative booting mechanisms (e.g., snapshot booting). In other words, users of PCs, servers, tablets, or smart phones keep their devices on or do not mind the booting time and traditional embedded devices of consumer electronics may use alternative booting mechanisms.


%
%

Therefore, it is difficult to optimize the launching time of emergent services: an application that shows the broadcast channel and services required to accept remote control signals and to control hardware accordingly. That is to say, the more the dependencies of services become intricate, the more emergent services have to be delayed. Even worse, the complicated dependency structure with non-determinism and dynamicity result in a boot time that varies among instances.

%
%

%

In this paper, we describe the BB, isolating the boot time-critical services, such as mount, socket, D-Bus, and topmost services with \textit{Booting Booster Group Isolator} to solve the increased dependency complexity by adding a new service. Moreover, the system architecture of the BB drastically lightens the existing init scheme \textit{systemd} to handle the boot time critical consumer electronics as well as server environments, to improve the boot time. The BB is based on \textit{systemd}, the standard init scheme of Tizen, which has been the OS of all Samsung smart TV products since 2015. The proposed approach does not require knowledge of the OS services by system developers who are working on the overall booting mechanisms (Linux kernel and \textit{systemd}), resolving one of the major disadvantages of dependency-based init schemes. Our system (Samsung Smart TV 2015 models) is orthogonal to prior mechanisms; i.e., the suspend-to-RAM technique is still additionally applied in high-end models.

%% file: subsection/050_design_implement.tex
\begin{figure}
\centering
\includegraphics[width=0.98\columnwidth,height=2.5in]{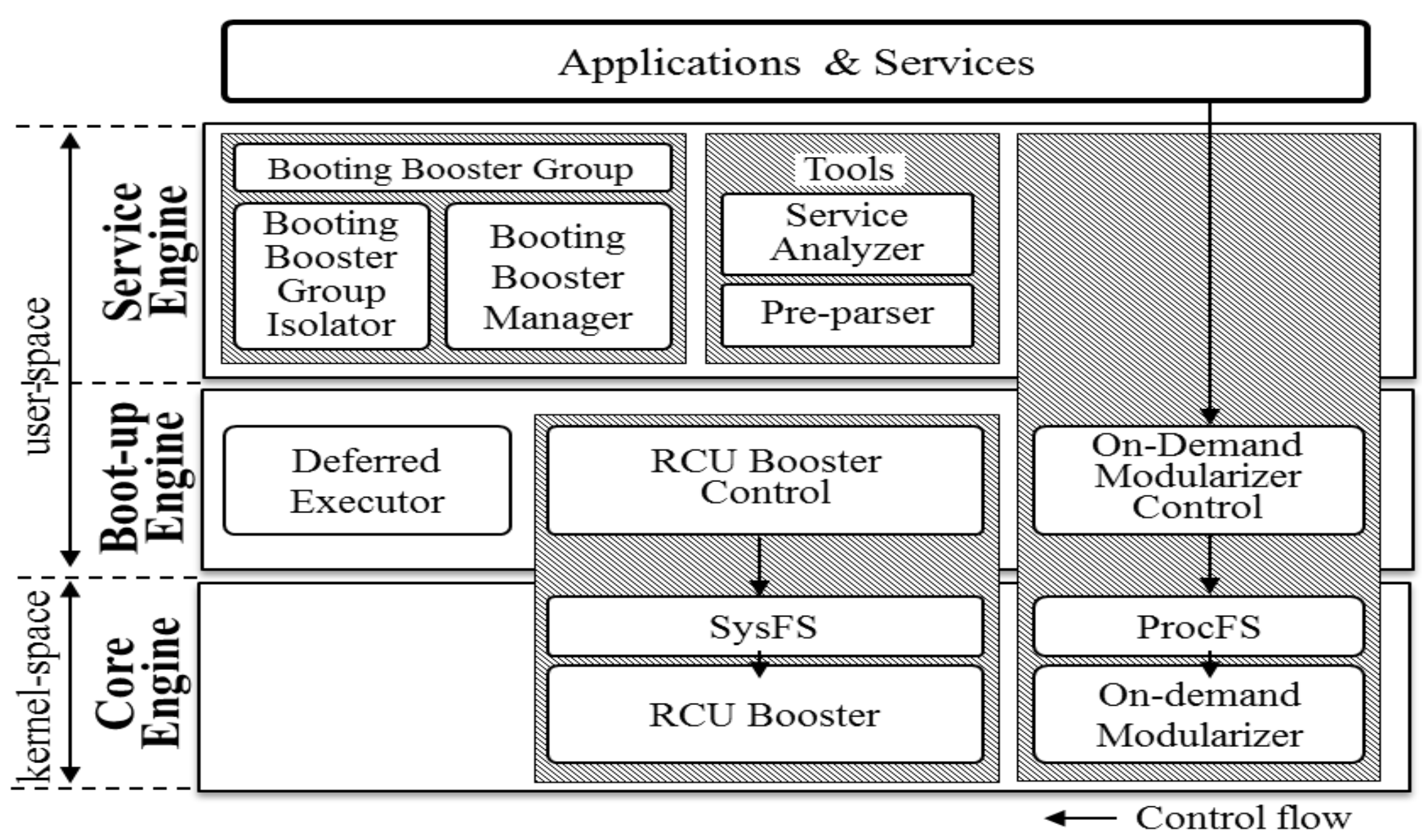}
\caption{Architecture of the Booting Booster (BB)}
\label{fig:quickboot-design}
\end{figure}

\begin{figure*}
\centering
\includegraphics[width=2.1\columnwidth,height=4.5in]{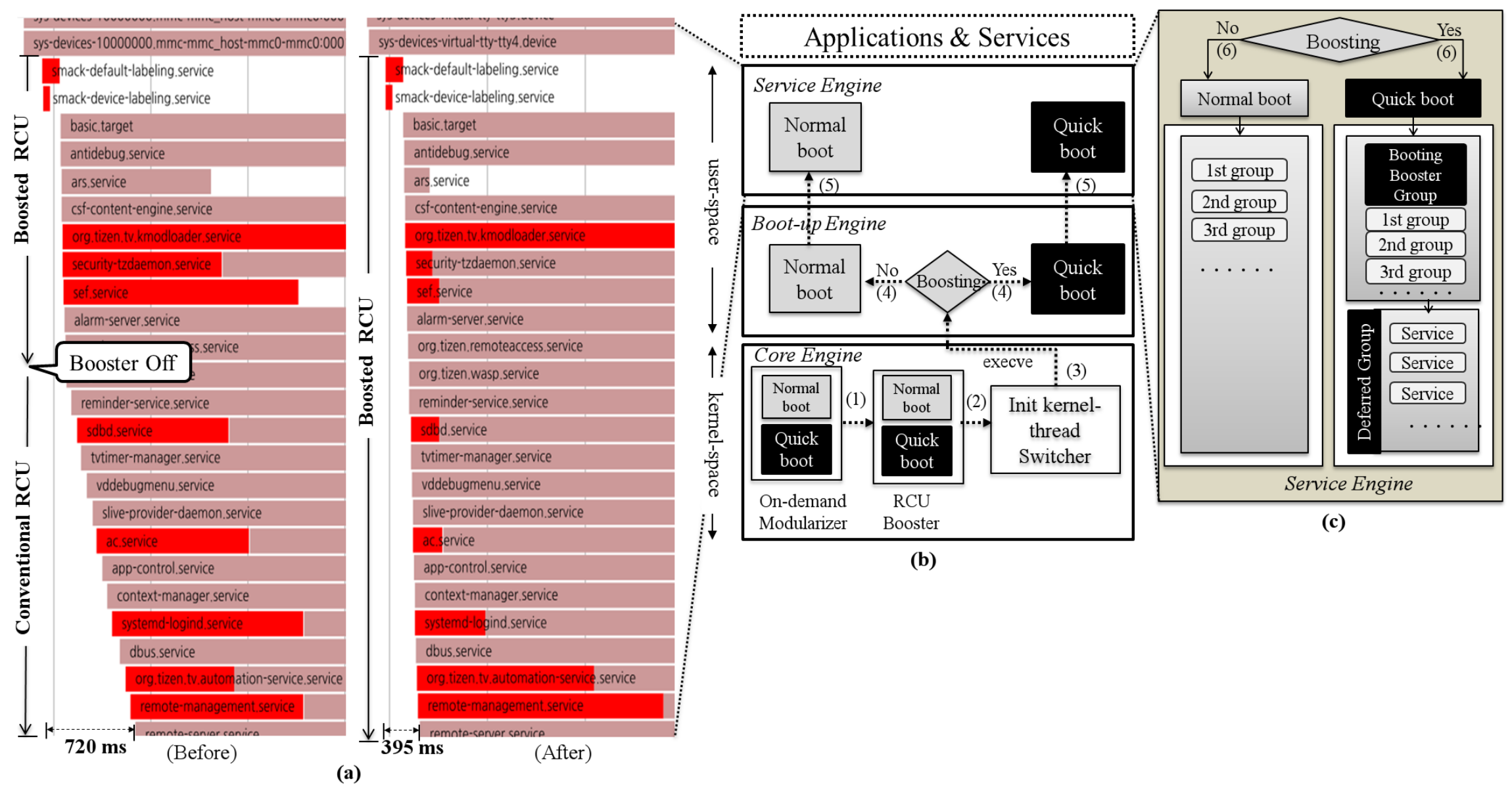}
\caption{Operation flow of the proposed system; (a) Graphs of systemd-bootchart showing the effects of RCU Booster, (b) Sequential operation flow of three engines in \textit{``Booting Booster (BB)''}, and (c)  \textit{``BB Group''} in the service engine. The boot-up and service engines are implemented by enhancing the existing processes of \textit{systemd}.}
\label{fig:quickboot-operation-flow}
\end{figure*}

%
%
%

In this section, we describe how the system is implemented for faster booting. Figure~\ref{fig:quickboot-design} shows the overall architecture of the proposed system, \textbf{Booting Booster (BB)}, implemented in Tizen. BB consists of three engines: the \textit{Core Engine}, \textit{Boot-up Engine}, and
\textit{Service Engine}. A detailed analysis of the time saved by each feature is described in \S\ref{S_evaluation}.
In abstract, our work boosts the booting sequences of Linux systems as follows: a) identify the BB Group and provide an isolated environment for the BB Group; b) defer modulation of built-in kernel features, defer execution of non-booting critical tasks in \textit{systemd}, and defer launches of non-booting critical tasks; and c) increase parallelism enabled by a boosted RCU and deferred tasks.

\subsection{Core Engine}\label{SS_core_engine}

%
%
%

The \textbf{Core Engine} consists of kernel-space BB components. \textit{On-demand Modularizer} modularizes built-in kernel components, which defers and concurrently starts subsystems not required to start the init scheme later. It defers the initialization of non-critical built-in kernel modules until completing the primary boot sequence, to run the built-in kernel modules  at the starting time of device-related user applications such as Universal Serial Bus (USB) after booting. By introducing deferred initialization approach based on \textvtt{initcall} \cite{initcall} with the built-in kernel module without the external kernel module, we drastically reduced the number of system calls (e.g. open, read, and close) required to load many external modules into volatile memory. \textit{Core Engine} shortens the time to begin user processes by initializing only the required size of memory and defers initializing the remaining area, which may take too much time with modern large-memory computing devices. \textit{RCU Booster} selectively boosts the \textit{Read-Copy-Update (RCU)} \cite{paul07_rcu_fundamental,paul15_locking_performance,paul98_rcu_concurrency}, which is a synchronization mechanism usually used to protect shared variables between kernel components. RCU is especially efficient for reading shared variables and it is used with extreme frequency while booting, which in turn, becomes a major bottleneck for booting. Bottleneck of RCU is also suggested by the experimental results in \S\ref{SS_bootup_engine}; i.e., the RCU overhead of booting time is greatly reduced by enabling \textit{RCU Booster} of \textit{Core Engine}.

\begin{algorithm}
\algsetup{linenosize=\tiny}
\caption{Conventional RCU synchronization algorithm~\cite{ticket_spinlocks,paul15_locking_performance}. Processor is busy doing nothing until lock is granted, wasting CPU cycles.}
\label{conventional-rcu}
\SetAlgoLined
\mbox{Synchronize\_RCU( ):}
\Begin{
Initialize RCU head on stack for dynamic init.\;
Wait for completion (Task is uninterruptible)\;
\eIf{!done}{
Add task to wait queue tail\;
\While{until action is done and timeout}{
Do spin lock irq function (wait-lock)\;
Set current state\;
Do spin unlock irq function (wait-lock)\;
}
}{
 Return timeout\;
 }
Destroy RCU head on stack for debug objects\;
}
\end{algorithm}

\begin{algorithm}
\algsetup{linenosize=\tiny}
\caption{Boosted RCU synchronization algorithm with a blocking lock improves performance in the contended case for accelerating boot-up time. This incurs greater CPU utilization due to process context switch and scheduling cost.}
\label{boosted-rcu}
\SetAlgoLined
\mbox{Synchronize\_RCU( ):}
\Begin{
SMP memory barrier\;
Snapshot accessed by other CPUs\;
SMP memory barrier\;
\While{mutex lock not locked}{
Try mutex lock\;
}
Force all RCU readers onto task lists\;
Do synchronized scheduling\;
SMP memory barrier\;
Compare snapshot\;
Do mutex unlock\; 
SMP memory barrier\;
}
\end{algorithm}

RCU is boosted selectively because the conventional mechanism (Algorithm \ref{conventional-rcu}) performs better if the number of threads using RCU is 0 or 1; it is normally 0 or 1 after booting completion. The conventional RCU mechanism protects critical sections of the mechanism by means of the ticket spinlock mechanism (since Linux 2.6.25) \cite{ticket_spinlocks}, which may be inefficient if there are multiple threads spinning to acquire the lock of a critical section at boot time. The implementation of \textit{RCU Booster} (Algorithm \ref{boosted-rcu}) replaces the need for ticket spinlocks by employing memory barriers and mutexes, allowing waiting threads not to spin but to sleep, which in turn, releases the CPU for other threads. However, as mentioned above, \textit{RCU Booster} has higher CPU load compared to the conventional if few threads are trying to acquire the lock simultaneously, so we use RCU Booster selectively.

Figure \ref{fig:quickboot-operation-flow}(b) shows the sequential operation flow of our proposed system. As the kernel becomes ready to start the first user process, which is the init scheme, \textit{systemd}, \textit{Core Engine} in the kernel starts its sub-components except for the memory initialization, which is executed earlier, before loading the first user process. More specifically,  \textit{On-demand Modularizer} generates built-in kernel modules that are decided to be deferred, and \textit{RCU Booster} is initialized.

\subsection{Boot-up Engine}\label{SS_bootup_engine}

%
%
The \textbf{Bootup Engine} consists of user-space BB components to run its own startup service. It is located in \textit{systemd}, which is the de facto standard init scheme of Linux. \textit{Boot-up Engine} is in charge of faster initialization of \textit{systemd} and enabling the facilities of kernel-space \textit{Core Engine} (e.g., ``(b) Boot-up Engine \#1'' and ``(c) Boot-up Engine \#2'' in Figure~\ref{fig:evaluation-all}) as the first task of \textit{systemd} before launching user-space services. \textit{RCU Booster Control} enables and disables \textit{RCU Booster} as a user-space agent. \textit{Deferred Executor} accelerates the initialization of the init scheme by deferring and executing in parallel init scheme sub-modules that are not required to start OS services. \textit{On-demand Modularizer Control} lets a component modularized by \textit{On-demand Modularizer} start as a user-space on-demand manager when the component is required by an application or an OS service.

%

%

As \textit{systemd} starts, Boot-up Engine is started as the first module of \textit{systemd}. \textit{RCU Booster Control} enables \textit{RCU Booster} as soon as Boot-up Engine is started and disables usually when the booting sequence is completed. The condition to disable \textit{RCU Booster} is configurable based on the decision of administrators; i.e., whether we can afford the CPU overhead to accelerate RCU functions, which are heavily used for booting. Figure ~\ref{fig:quickboot-operation-flow}(a), drawn with systemd-bootchart \cite{systemd_bootchart}, shows that more tasks are quickly launched in parallel at booting with \textit{RCU Booster}. In the figure, the x-axis shows time and the y-axis shows OS services being launched from the top to the bottom. The boosted case shows earlier launching of a greater number of tasks; i.e., services in the bottom start earlier (nearer to the origin of the x-axis).

%
%

With \textit{Deferred Executor}, \textit{Boot-up Engine} defers \textit{systemd} tasks not required to start launching OS services until the system recognizes booting completion; i.e., the user is watching broadcast channels. In this case, essential services are identified as the minimal OS services required to watch broadcast channels and respond to remote control input. Thus, \textit{systemd} may execute required tasks only and start launching OS services earlier. Deferred \textit{systemd} tasks are logging and setup procedures, including kernel modules, hostname, machine ID, loopback device, and directories with test and debug purposes. Enabling EXT4 journal mode of the root file system is deferred as well because we virtually are read-only while booting and we can remount the root file system is writable journal mode later as a deferred task.

\subsection{Service Engine}\label{SS_service_engine}

%
%

The \textbf{Service Engine} consists of user-space BB components to control booting-critical and non-critical services. It has tools for system administrators and \textit{Booting Booster Group Isolator} and \textit{Booting Booster Manager}, which are implemented in \textit{systemd} and accelerate the booting sequence when \textit{Service Engine} in \textit{systemd} is ready to start launching user-space services. As \textit{systemd} is ready to launch user-space services (\textit{Boot-up Engine} is fully up and running well before this) concurrently, \textit{Booting Booster Group Isolator} identifies a group of booting critical processes, \textit{``Booting Booster (BB) Group''} in Figure ~\ref{fig:quickboot-operation-flow}(c), which are OS services required for a user to recognize that the system is ready to use; i.e., services required to show a TV channel, to handle remote controller signals, and by the first application launched. One motivation for isolating the boot sequence after dependency analysis is to avoid developers prioritizing their own services to the detriment of the system as a whole. Note that developers do not play the same games with the critical path by creating false dependencies. In a 2015 Samsung smart TV model, there were seven services (i.e., mount, socket, dbus, tuner, hdmi, demux, and fasttv) in the BB group. \textit{Booting Booster Manager} launches processes of the BB Group and prioritizes and manages processes of the group to complete booting quickly.

%
%

Tools in Service Engine help system administrators optimize the booting procedures: \textit{Service Analyzer} and \textit{Pre-parser}. \textit{Service Analyzer} investigates the relations between services by reading the configuration files of software packages and reports incorrect relations (i.e., circular dependencies and contradicting requirements) based on call-graph generators: Codeviz \cite{codeviz} and Graphviz \cite{graphviz}. \textit{Pre-parser} reduces the boot-time overhead of parsing service configuration files, which are text files written by hundreds of services launched by \textit{systemd}. Pre-parser parses such service configuration files beforehand and allows \textit{systemd} to read pre-parsed data and to skip reading and parsing the configuration files at boot time.

A configuration file  of a software package \cite{systemd_unit,systemd_official_freedesktop} is shown in Listing \ref{configuration-file}. The ``Before=socket.service'' expression means that the \textvtt{socket.service} unit cannot be started until the \textvtt{Myapp.service} is activated. If they are activated at the same time at boot time, a start time of each service is decided by the dependency relationship between the two service units.  There are three types of unit to decide a starting time of the service.  First, ``Type=Simple'' means that the latter service starts as soon as the former service starts. This type can be used to run independent tasks between the services. Second, ``Type=Forking'' means  that  socket.service can be started as soon as  ``ExecStart='' statement's fork system call executes.  Finally,  ``Type=oneshot'' means that  \textvtt{socket.service} can be executed as soon as ``ExecStart='' statement completes with fork system call. The ``WantedBy=multi-user.target'' statement means that the \textvtt{Myapp.service} unit belongs to the \textvtt{multi-user.target group}. 

\lstset{
language=C,
basicstyle=\small,
numbers=left,
numberstyle=\tiny,
frame=tb,
columns=fullflexible,
showstringspaces=false,
moredelim=[is][keywordstyle]{(@@}{@@)},
moredelim=[is][\underbar]{(*}{*)}
}

\begin{lstlisting}[label=configuration-file,caption=A configuration file of \textvtt{Myapp.service} to define the dependency relationship among the services,float=t]
(@@[Myapp.service]@@)
(@@[Unit]@@)
Description=Summarized explanation of Myapp.service
(*Before=socket.service*)
(@@[Service]@@)
Type=oneshot
ExecStart=/usr/bin/myapp-service-daemon
(@@[Install]@@)
WantedBy=multi-user.target
\end{lstlisting}

%
%

When \textit{Service Engine} starts, \textit{Booting Booster Group Isolator} identifies  \textit{BB Group} services that are critical to booting completion. \textit{Booting Booster Group Isolator} identifies such services by analyzing relations spanning from the dependencies of the definition of boot completion. The isolated \textit{BB Group} allows the corresponding services to ignore services not in the group and dependencies or priority requirements defined as out of the group. With \textit{BB Group}, system administrators can maintain a consistent booting time with on-going development of other OS services and applications and focus on booting-critical tasks only. For example, even if messenger services declare that they are to be started before a broadcast signal-handling service (a booting-critical task), launching the broadcast signal-handling service is not affected by messenger services; as long as the broadcast signal-handling service and its explicitly required services do not explicitly require messenger services.

%
%

Then, \textit{Booting Booster Manager} starts launching processes in the group. At this point, processes not in the BB Group are being launched by \textit{systemd} as well as those in the group. \textit{Booting Booster Manager} prioritizes processes in the \textit{BB Group} for faster booting. As a result, processes not in the group are deferred if computing resources are not available.

%% file: subsection/060_evaluation.tex
We show the experimental results of a Samsung UHD Smart TV model (\textvtt{UE48H6200}), shipped in 2014. Like its successors, all 2015 models of Samsung Smart TVs, \textvtt{UE48H6200} runs Tizen 2.3 TV profile \cite{opensource_samsung} and Linux 3.10 kernel along with \textit{systemd} v208 init. \textvtt{UE48H6200} has an application processor with four Cortex A9 CPU cores, 1~GiB DRAM, a GPU, and video and audio processing units along with 8~GiB eMMC flash storage, network interfaces, and a 48-inch UHD display panel. Note that the performance of the eMMC flash storage is not comparable with SSDs, but is comparable with consumer HDDs. The eMMC of \textvtt{UE48H6200} has a sequential read performance of 117~MiB/s and a random read performance of 37~MiB/s A consumer SSD, the Samsung SSD 850 Evo 500~GB, has a sequential read performance of 515~MiB/s and a random read performance of 379~MiB/s. A consumer HDD, the Seagate Barracuda 3TB (\textvtt{ST3000DM001}, released in 2011), has a sequential read performance of 165 MB/s and a random read performance of 65 MB/s.

%
%
%
%

%

The source code of the system is available at \url{http://opensource.samsung.com/} with the model name ``U***H62**''. Although we present experimental results of a single 2014 smart TV model, UE48H6200, the presented work, BB, is widely applied to many 2014 models of Samsung Smart TV that have the same hardware platform. Moreover, all 2015 Samsung Smart TV models use the same version of Tizen, kernel, and BB presented in this paper and satisfy the performance requirements as well.

The de facto standard init scheme of Linux, \textit{systemd}, is used for most smart TV sets as well as for desktop and server systems. The global smart TV market is shared mainly (over a half) by three major manufacturers: Samsung (Tizen), LG (WebOS), and Sony (CE Linux\footnote{Sony is moving to Android TV from their in-house OS \cite{Sony_Android_TV}.}) \cite{tv_market_share_2014,tv_market_share_2015Q1,tv_market_share_2015Q2}. Because all three major smart TV OSs use a Linux kernel and  \textit{systemd}, the presented work, BB, can be easily ported to the other two OSs. In addition to the smart TV sets, BB has been applied to diverse devices, including mobile phones (Samsung Z1 and Z3, since 2015), wearable devices (Gear series, since 2014), digital cameras (NX series after NX300, since 2013), and other home appliances (air conditioners, refrigerators, and robotic vacuum cleaners, since 2015). Therefore, BB can be seamlessly and easily applied to a wide range of consumer electronics.

\begin{figure*}
\centering
\includegraphics[width=1.0\textwidth,height=5.0in]{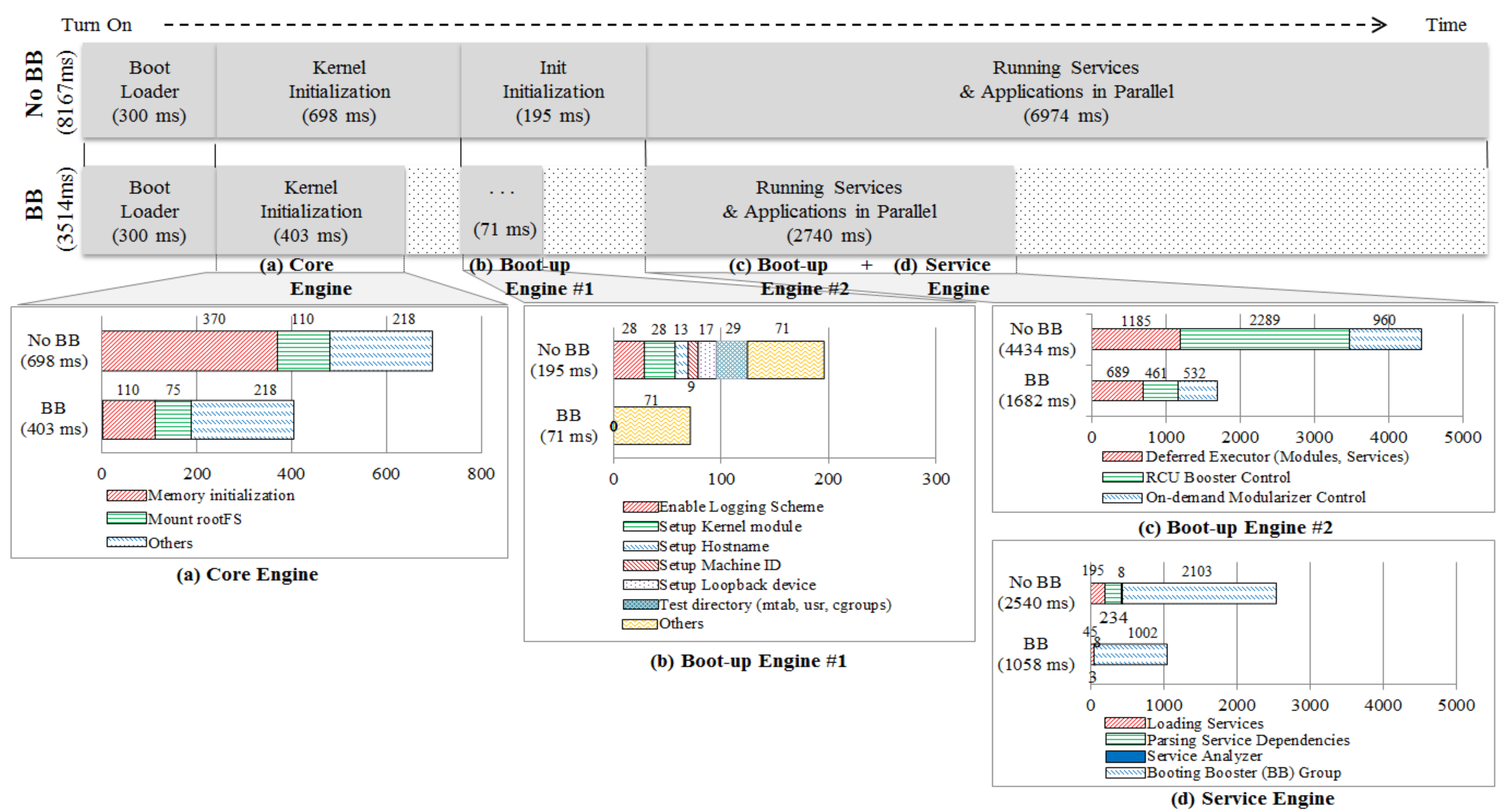}
\caption{Comparison of the two experimental sets: the conventional (No BB) and BB and analysis of each booting step}
\label{fig:evaluation-all}
\end{figure*}

\subsection{Experimental Results}

%

Figure~\ref{fig:evaluation-all} shows the system-wide experimental results of booting time for both the conventional and BB mechanisms. The time is measured using a high-precision hardware timer with nanosecond precision integrated in the application process, which begins with the power-on signal. The experiments suggest that BB reduced the booting latency by $\sim$57\% from 8.1~s to 3.5~s.

%

For the analysis, we measure the latency of three major steps of the whole booting sequence, as shown in Figure~\ref{fig:evaluation-all}. We also show the comparative analysis of the conventional and BB for sub-steps for each step divided by the three major steps and the three engines of BB. Because the second engine, \textit{Boot-up Engine}, is executed in the later two major steps, the analysis of \textit{Boot-up Engine} is divided into two parts, as shown in the figure. In the following list, we describe the reduction in latency due to each specific method of the proposed mechanism, as described in Figure~\ref{fig:evaluation-all}.

\begin{enumerate}
\item \textbf{Kernel initialization (a):} \textit{Core Engine} has reduced kernel
initialization latency from 698~ms to 403~ms with memory initialization
(370$\rightarrow$110 ms) and root filesystem mounting (110$\rightarrow$75 ms),
which are deferred until boot completion.
\item \textbf{Init initialization (b):} The earlier part of \textit{Boot-up Engine}
initializes init. \textit{Boot-up Engine} has reduced init initialization from 195~ms to
71~ms by deferring various tasks that are not crucial for booting completion.
The deferred tasks are ``enable logging scheme'' (28 ms), ``setup kernel module'' (28 ms),
``setup hostname'' (13 ms), ``setup machine ID'' (9 ms), ``setup loopback device'' (17 ms),
and ``test directory'' (29 ms) from left to right in Figure~\ref{fig:evaluation-all}(b).
In total, we have deferred 124~ms worth of tasks until boot completion without visible overhead.
\item \textbf{Running services \& applications in parallel (c)+(d):}
The later parts, \textit{Boot-up Engine} and \textit{Service Engine} are included in
the phase of running services \& applications in parallel.
The effect of the kernel-space \textit{RCU Booster} and the user-space \textit{RCU Booster Control} is represented
by \textit{RCU Booster} in Figure~\ref{fig:evaluation-all}(c): 2289~ms$\rightarrow$461~ms.
\textit{Deferred Executor} has saved 496~ms and \textit{On-demand Modularizer} along with
\textit{On-demand Modularizer Control} has saved 428~ms.
Figure~\ref{fig:evaluation-all}(d) shows that \textit{Pre-parser} has saved 150~ms for
``loading services'' and 231~ms for ``parsing service dependencies'' and
\textit{Booting Booster Group Isolator} and \textit{Manager} have saved 1101~ms
by isolating booting critical tasks with \textit{BB Group}.
\end{enumerate}


Overall, the results suggest that task isolation with \textit{Booting Booster Group Isolator}
and \textit{Booting Booster Manager} (saves 1101~ms) and synchronization mechanism optimization
with \textit{RCU Booster} and \textit{RCU Booster Control} (saves 1828~ms) drastically
reduce the booting latency.

\subsection{How Booting Booster Group Isolation Works}\label{SS_eval_isolation}

\begin{figure*}
\centering
\includegraphics[width=0.99\textwidth,height=5.5in]{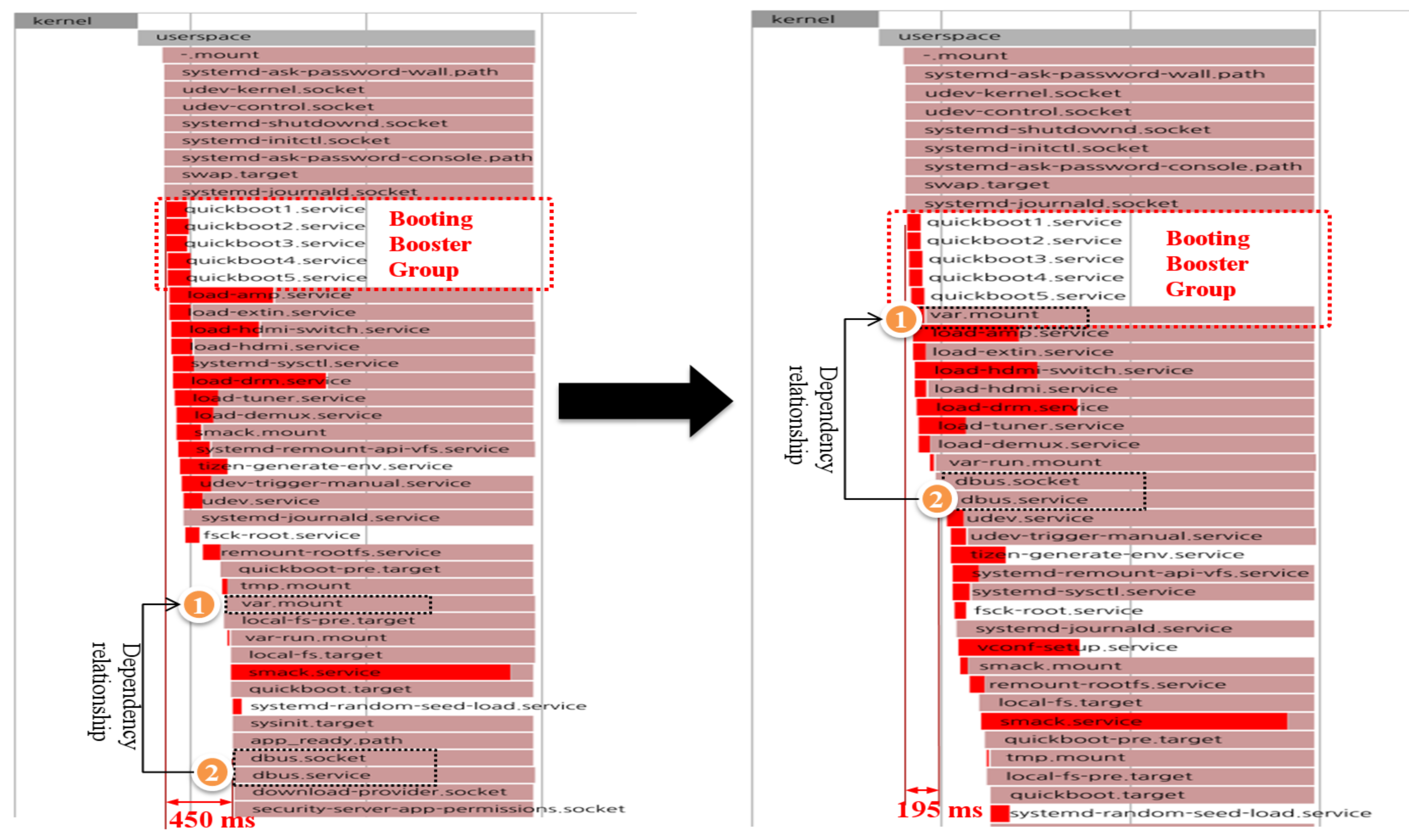}
\caption{Graphs of systemd-bootchart showing the effects of \textit{BB Group}
by adding \textvtt{var.mount} into \textit{BB Group} to advance the launching time of \textvtt{dbus.service}}
\label{fig:quickboot-eval-with-dbus}
\end{figure*}

Figure~\ref{fig:quickboot-eval-with-dbus} shows the experimental results
of the conventional and a partial execution of \textit{Booting Booster Group Isolator},
focused on the launching time of ``dbus.service'' (standard IPC mechanism) and
``var.mount'' (mount the \textvtt{/var} directory).
Both services are booting critical and ``dbus.service'' depends on the completion of
``var.mount''.
Because ``dbus.service'' is an essential service for virtually all
Tizen services and applications, it is recommended to be launched as early as possible.
The results suggest that the launching time of ``dbus.service'' is significantly
advanced by isolating ``var.mount'': 450~ms compared to 195~ms.

The left side of Figure~\ref{fig:quickboot-eval-with-dbus} shows a comparison of the booting sequence
with the conventional.
The right side shows that the booting sequence  is boosted compared to the conventional.
In the figure, \textcircled{1} is ``var.mount'' and \textcircled{2} is ``dbus.service''.
For the boosted case, we have manually added ``var.mount'' into the isolated BB group
without other booting--critical services
that would have been added to the group when \textit{Booting Booster Group Isolator} is fully enabled,
which automatically identifies booting--critical services by recognizing \textit{BB Group} at booting time and executes \textit{BB Group} as a topmost job. 
To evaluate the effect focused on ``var.mount'' and its successor, ``dbus.service'',
we did not enable ''dbus.service'' with \textit{Booting Booster Group Isolator}, 
but manually declared ``var.mount'' as a booting critical task.

Although, system administrators instruct developers not to do so, service and application developers have added ordering dependencies between their own services (about a dozen in the final release) and ``var.mount'' so that their services may be launched as soon as possible to make them appear more optimized. Besides, ``var.mount'' is not the only case; some other booting--critical tasks automatically identified by BB suffer from the same issue. In such environments, we can successfully launch essential services in advance and complete booting in time by isolating booting--critical tasks from other processes.


\subsection{Performance Trade Off}

%

\textit{RCU Booster} consumes more CPU resources compared to the conventional RCU mechanism if there is no thread competing to synchronize an RCU. This is because \textit{RCU Booster} uses pre-emptible mutexes instead of spinlocks to use the CPU more efficiently if many threads are attempting RCUs during booting. \textit{RCU Booster} provides the dynamic control interface using the \textvtt{sysfs} filesystem \cite{mochel05_sysfs} to the user space so that \textit{RCU Booster Control} of the user-space may enable and disable \textit{RCU Booster} dynamically. In general run time (except for cases such as booting), there are usually no two threads competing to synchronize a single RCU, and in such cases a spinlock is cheaper than a mutex.

%
%

BB depends on deferring tasks not crucial to booting. Deferred tasks are executed after boot completion, while deferred tasks should be executed together with other booting tasks. Besides, deferring tasks that were executed unconditionally in an earlier phase (i.e., during \textit{systemd} launch) allows the deferred tasks to be executed in parallel. However, deferring such system services makes applications dependent on them incur additional launch delay, although such applications do not determine the booting time. Fortunately, once an application triggers a deferred task to start, the deferred task no longer incurs an additional delay for following application launches and an application usually does not depend on more than two such deferred tasks. The experiments suggest that the performance overhead of deferring tasks is negligible: less than 15 ms on average and the standard deviation less than 1.5 \% for applications that depend on deferred service tasks.

%% file: subsection/070_discussion.tex
\textbf{Dependency on open source packages.} Consumer electronics software platforms may include a vast number of open-source software packages; Tizen has hundreds. Manufacturers use open-source packages because they reduce the cost of development and maintenance, allow employment of emerging technologies with less effort, and allow external developers to contribute. On the other hand, employing various open-source software packages increases the dependency complexity.

The experimental results with task isolation suggest that disconnecting dependencies between OS components might be crucial for optimization of booting speed. Software fragmentation usually occurs if we modify such open-source packages to cut the dependency relations. Most \textit{upstream} packages evolve independently from a single software platform, which is merely one of the users of the \textit{upstream} open-source software packages. Thus, we need either to merge our modifications into the \textit{upstream} quickly before the \textit{upstream} evolves too much or to keep local \textit{downstream} forks updating with \textit{upstreams} regularly. In the long run, the former is surely the better method; However, with hundreds of such packages being modified by different divisions and device categories, it is often too difficult to achieve.
%

\vspace{5pt}

\vspace{5pt}

\textbf{Usage and effectiveness of BB Group.} We proposed isolation of booting-critical processes, which achieves shorter booting time with \textit{BB Group}. The proposed technique does not require modification of the existing software packages. It only requires listing the applications that define booting completion; i.e., TV broadcast apps in televisions. Isolation allows non-booting critical tasks to be deferred and the CPU to load booting-critical tasks defined by required completion time of the applications. Without isolation and deferring, many booting-critical tasks wait for CPU cores while these are made use of by non-critical tasks. \\

\textbf{Tackle dependencies directly.} Ideally, minimum dependency relations between processes would be maintained by identifying unnecessary dependencies or modifying software to cut dependencies. Reducing dependency relations allows the system to further utilize parallelism, to find a shorter path to booting completion, and even to identify unnecessary software packages. However, removing dependency declarations incorrectly may disable the booting sequence; or even worse, it may jeopardize in a non-deterministic way hidden to product testers, a nightmare for manufacturers. As mentioned above, it is extremely difficult for system administrators to understand and update other software packages simply because there are too many in modern software platforms; e.g., $\sim$1000 packages in Tizen, excluding third-party apps and vendor- and operator-specific packages. Besides, some developers tend to declare excessive dependencies to feel safer or inappropriate orderings to improve the performance of their own package only. Therefore, to identify the minimal dependency structure, we are virtually forced to ignore what they have declared by experimenting with all possible launching sequences.

%
%

We once attempted to audit every additional dependencies or ordering declarations. In practice, the volume of changes was too great to be handled by a few administrators and virtually all incoming software packages were totally unfamiliar to the administrators. Besides, many developers did not care even to inform that they are adding a new software package to the system, not even a new dependency between packages. Making things even worse, such changes were made daily even when the product shipment date was very near; sometimes, a change to the OS is made even after the products have shipped.

%
%

Fortunately, in the case of most consumer electronics, booting completion is defined by the completion of a few user processes, not all user processes that are launched by the OS at booting time. Even better, such few processes require only a few OS services, which make it possible for few administrators to tackle inter-service dependencies as long as the processes are completely isolated by \textit{BB Group}.

%
%

We have not tackled dependencies directly, yet, because we could achieve the required booting time without analyzing other unfamiliar software components. Although it is not recommended, if the size of \textit{BB Group} grows (and surely will grow in a few years), an automated mechanism will be required to verify dependency declarations to remove or add dependencies. Note that such a mechanism is not trivial; source codes of some packages are not available even to manufacturers and some dependencies are not directly shown (dependencies based on the availability of a file path or the value of a file).

\vspace{5pt}

\textbf{Pre-parser, pre-link, and pre-fork} reduce computation resources (time and memory) with pre-processing mechanisms. \textit{Pre-parser} of BB reduces the CPU time to parse all service declarations (\textit{systemd} unit files) at boot time by parsing the whole data beforehand and keeping the parsed data to avoid parsing it for each instance of booting.

Pre-link and pre-fork are traditional mechanisms of saving resources consumed by launching of user processes. However, for the processes in \textit{BB Group}, we do not apply pre-link or pre-fork. This is because the two traditional mechanisms may incur a security issue and more overhead without performance benefit for the group. To further optimize programs in \textit{BB Group}, we statically built the processes in the group, which completely removes overheads incurred by dynamic linking. Besides, there are usually no preceding processes with the same library for the processes in the group because it is at a very early stage of the booting sequence. Thus, pre-link for \textit{BB Group} shows no benefit, although processes not in the group might be launched faster if processes in the group use pre-link. Note that optimizing non-booting-critical processes is not a concern for optimizing booting performance.

Pre-fork is also not beneficial for processes in the group because pre-fork requires heavy performance overhead for starting the pre-fork mechanism itself. Because the \textit{BB Group} is executed in a very short time with few processes, the benefit (reduced time to create user processes) of pre-fork does not exceed the overhead (increased time to pre-launch user processes).


%% file: subsection/080_conclusion.tex
%
%

With the introduction of multi-core architecture, init schemes have evolved to launch processes in parallel during booting. The increasing complexity of software platforms results in slower booting of embedded devices with modern OSs and rich services. The booting time is an extremely significant performance metric in many consumer electronics devices, such as televisions and digital cameras. We have successfully reduced the booting time of Tizen-based consumer electronics, smart TVs, from 8.1~s to 3.5~s using mechanisms that can be applied generally to Linux systems with a de facto standard init scheme, \textit{systemd}.

%
%
%

The proposed mechanism is available as open-source software at \url{http://opensource.samsung.com/} and has been adopted widely in all 2015 Samsung Smart TV models globally. Also, the proposed mechanism will be released with later versions of the Tizen TV profile at \url{http://tizen.org/}.

%% file: subsection/090_acknowledgment.tex
We would like to thank the anonymous reviewers and our shepherd Kevin Elphinstone for their valuable feedback on earlier drafts of this paper.
We also gratefully appreciate Sangjung Woo and Jae-Young Hwang of Samsung Electronics
for the insightful and extensive discussions and comments.
This work is supported in part by Frontier CS Research 2015 and Tizen 2015
through Software R\&D Center of Samsung Electronics.

%% file: subsection/100_reference.tex

\bibliographystyle{abbrv}
\bibliography{main}

%
%
%
%